\documentclass[sigconf,natbib=false]{acmart}
\AtBeginDocument{%
  }

\copyrightyear{2023} 
\acmYear{2023} 
\setcopyright{acmlicensed}\acmConference[SOICT 2023]{The 12th International
Symposium on Information and Communication Technology}{December 7--8,
2023}{Ho Chi Minh, Vietnam}
\acmBooktitle{The 12th International Symposium on Information and
Communication Technology (SOICT 2023), December 7--8, 2023, Ho Chi Minh,
Vietnam}
\acmPrice{15.00}
\acmDOI{10.1145/3628797.3628987}
\acmISBN{979-8-4007-0891-6/23/12}

\usepackage[numbers]{natbib}

\begin{document}

\title{IncepSE: Leveraging InceptionTime's performance with Squeeze and Excitation mechanism in ECG analysis}

\author{Tue M. Cao}
\email{tue.cm@vinuni.edu.vn}
\affiliation{%
  \institution{School of Information and Communication Technology, Hanoi University of Science and Technology}
  \institution{VinUni-Illinois Smart Health Center, VinUniversity}
  \city{Hanoi}
  \country{Vietnam}
}

\author{Nhat H. Tran}
\email{nhat.th2@vinuni.edu.vn}
\affiliation{%
  \institution{School of Information and Communication Technology, Hanoi University of Science and Technology}
  \institution{VinUni-Illinois Smart Health Center, VinUniversity}
  \city{Hanoi}
  \country{Vietnam}
}

\author{Le P. Nguyen}
\email{lenp@soict.hust.edu.vn}
\affiliation{%
  \institution{Hanoi University of Science and Technology}
  \city{Hanoi}
  \country{Vietnam}
}

\author{Hung T. Nguyen}
\email{hungnt@soict.hust.edu.vn}
\affiliation{%
  \institution{Hanoi University of Science and Technology}
  \city{Hanoi}
  \country{Vietnam}
}

\author{Hieu H. Pham}
\authornotemark[1]
\thanks{* Corresponding author.}
\email{hieu.ph@vinuni.edu.vn}
\affiliation{%
  \institution{VinUni-Illinois Smart Health Center, VinUniversity}
  \city{Hanoi}
  \country{Vietnam}
}

\renewcommand{\shortauthors}{Tue et al.}

\pagenumbering{arabic}
\begin{abstract}
  Our study focuses on the potential for modifications of Inception-like architecture within the electrocardiogram (ECG) domain. To this end, we introduce IncepSE, a novel network characterized by strategic architectural incorporation that leverages the strengths of both InceptionTime and channel attention mechanisms. Furthermore, we propose a training setup that employs stabilization techniques that are aimed at tackling the formidable challenges of severe imbalance dataset PTB-XL and gradient corruption. By this means, we manage to set a new height for deep learning model in a supervised learning manner across the majority of tasks. Our model consistently surpasses InceptionTime by substantial margins compared to other state-of-the-arts in this domain, noticeably 0.013 AUROC score improvement in the "all" task, while also mitigating the inherent dataset fluctuations during training.
\end{abstract}

\ccsdesc[300]{Applied computing~Life and medical sciences}
\ccsdesc[500]{Computing methodologies~Computer vision}

\keywords{Computer vision, ECG, Timeseries classification, Inception, Squeeze and Excitation}

\maketitle

\section{Introduction}
Cardiovascular disease is the leading cause of death worldwide \cite{allender2008european}. Timely intervention can prevent severe heart-related incidents, with the most crucial tool for detecting and diagnosing cardiac electrical irregularities being the electrocardiogram (ECG). With more personal ECG devices becoming affordable and widely available, ECG has become an essential tool for diagnosing cardiovascular diseases. The ECG serves as a noninvasive representation of the electrical activity of the heart measured using electrodes placed on the torso. The standard 12-lead ECG is widely used to diagnose a variety of cardiac arrhythmias such as atrial fibrillation and other cardiac anatomy abnormalities such as ventricular hypertrophy. With the rising of smart healthcare, automated ECG analysis has the potential to aid healthcare professionals in clinical settings and offer the general population a means of continuous monitoring through wearable devices. Therefore, improving accuracy and trustworthiness of automatic cardiovascular diseases emerges as a crucial concern.

In the past decades, there have been a large number of ECG classification methods proposed and developed. While traditional methods involve a number of  handcrafted features such as statistical, wavelet features and mathematical transformations, deep learning methods perform feature extraction via deep neural networks. Deep learning has achieved significant advancements in smart healthcare due to its ability to automatically extract abstract features automatically \cite{shenavarmasouleh2023deep}. Therefore, many deep learning-based approaches for ECG analysis have been proposed over the last decade, specifically relying on 1D-CNNs \cite{hong2020opportunities, STT-CNN5} or RNN-based \cite{Strodthoff:2020Deep,selfsupervisedECG} networks rather than 1D-transformers. Transformers fall short compared to their counterparts due to the periodic nature of ECG signal requires complex local feature extraction, while their strengths lie heavily in extracting global features. In contrast, the convolutional layer has shown incredible performances \cite{diagnosticsSE}\cite{Physionet1} with a powerful mechanism to capture local perturbations both temporally and spatially. One outstanding example is InceptionTime \cite{DBLP:journals/corr/abs-1909-04939}, which has been evaluated with competitive performance in many challenging tasks \cite{Strodthoff:2020Deep}\cite{eegeyenet}\cite{DBLP:journals/corr/abs-1909-04939}. However, effective as it is, ECG classification is still a difficult problem for 1D-CNN, with various tasks that are exceptionally hard to distinguish. Moreover, instability arises as a common problem while training since the datasets are often imbalanced in medical data, hence, deep learning models are susceptible to overfit and fluctuation during training. Thus, to address these challenges, we proposed an integration of Squeeze-and-Excitation which was proven to emphasize informative channels and suppress less relevant ones \cite{SENet}. We will show that this novel technique, combined with several adjustments in branches and skip-connections of InceptionTime overcomes the fluctuation and overfitting problem, leveraging the robustness and generality of the original architecture, reaching state-of-the-art performance in well-studied ECG tasks. Additionally, recent studies mostly focus on the architecture design aspect, while ignoring the importance of the training dynamics. This is a great hindrance to these architectures’ performance     due to the fact that ECG datasets are unbalanced and the trainings are unstable. To this end, a well-built training procedure must be defined to tackle the above limitation. We sum up our contributions as below:
\begin{itemize}
\item We propose IncepSE: an 1D-CNN architecture inspired by InceptionTime \cite{ismail2020inceptiontime} that utilizes multi-scale convolution filters and inception-like blocks as well as the squeeze and excitation mechanism \cite{SENet} to tackle ECG classification problem.

\item We design a training methodology with multiple additional components to tackle the problems of overfitting and gradient unstability. The proposed approach is tested on PTB-XL dataset \cite{wagner2020ptb}, and to the best of our knowledge, our method outperforms all others to date, reaching state-of-the-art performance on most of the classification tasks on this dataset.

\item We also conduct comprehensive ablation studies to evaluate the effect each component has on our model’s performance and show that each component plays an important part in the total performance gain of IncepSE.
\end{itemize}

The remainder of this paper is structured as follows: Section 2 offers an overview of other works relevant to our study. In Section 3, we present components of the proposed model and training procedure in detail. In Section 4, we outline the details of our experiments. In Section 5, we present the experimental results and conduct an ablation study. Finally, we give further discussions and conclusion in Section 6.

\section{Related Work}
Deep learning is a subset of machine learning employing multilayer neural networks to acquire data representations through several levels of abstraction. With great scalability, adaptability, and the ability to learn hierarchical representations of many types of data, deep learning has been the go-to choice in areas such as computer vision, speech recognition and autonomous systems \cite{sarker2021deep}. With the availability of large ECG datasets \cite{wagner2020ptb, liu2018open,reyna2021will} in recent years and significant advancements made in the field, deep learning has been applied to various tasks where ECG data is present.

Many different deep learning techniques have been employed for ECG data analysis. Algorithms based on recurrent neural networks (RNNs), including GRUs \cite{zhang2021mlbf} and LSTMs \cite{hochreiter1997long}, have demonstrated their effectiveness in ECG analysis by virtue of their capacity to extract temporal characteristics. Conversely, some approaches have utilized attention mechanisms \cite{9662723, zhang2021mlbf}, although these have been found to be less proficient in capturing local features, which are essential in ECG data due to their periodic nature. On the other hand, 1D-CNN based methods have emerged as highly effective \cite{wang2017time,schirrmeister2017deep,he2016deep}, as they can capture both low-level patterns (such as individual waveforms) and high-level patterns (like complex arrhythmias), aligning well with the periodic multi-channel nature of ECG data. Among those neural networks was InceptionTime \cite{ismail2020inceptiontime}, which consists of multi-scale convolution filters and inception-like blocks to capture diverse temporal patterns. However, InceptionTime was introduced as a general CNN architecture for time series classification, which comes at the cost of lowering its performance in some specific tasks. In our work, we propose a novel InceptionTime-based architecture combined with squeeze and excitation mechanism \cite{SENet}, which is shown to have significant improvements in the task of ECG classification.

\section{Methodology}

\subsection{Overview}
\begin{figure*}
    \centering
    \includegraphics[width=0.6\textwidth,keepaspectratio]{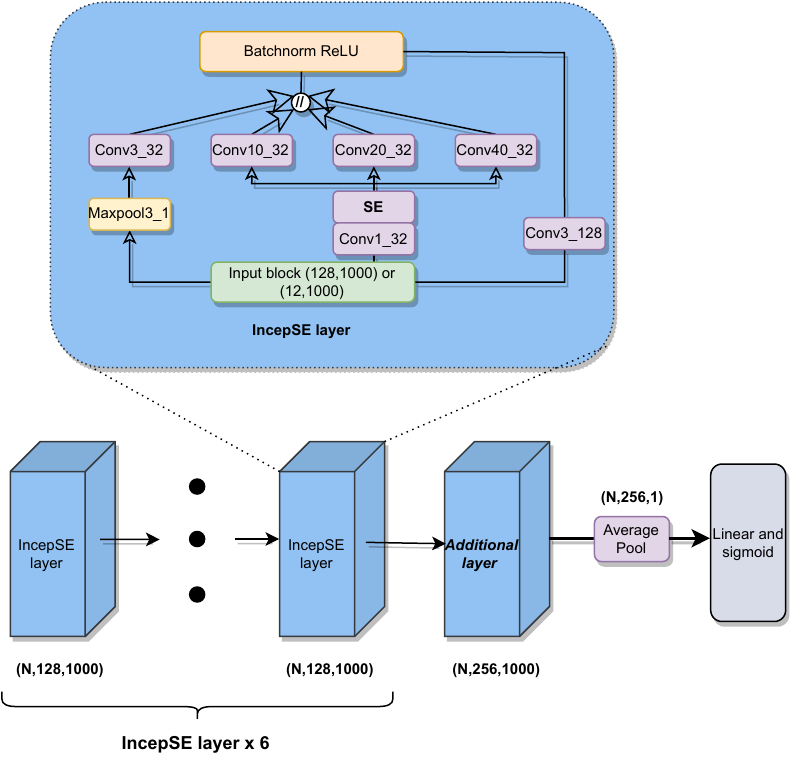}
    \caption{The overall architecture of the IncepSE model and each IncepSE layer}
    \label{fig:1}
\end{figure*}

The authors in \cite{Strodthoff:2020Deep} proposed a benchmark for the PTB-XL dataset, where they tested 7 different architectures that work well with timeseries data and claimed that InceptionTime \cite{DBLP:journals/corr/abs-1909-04939}, an 1D-CNN with multiple convolutional kernel sizes for diverse temporal feature extraction, gave outstanding performance across different classification tasks in the dataset. Inspired by InceptionTime as well as the effective squeeze and excitation mechanism (SE), we present IncepSE, harnessing the combined power of these two approaches. Our model is made up of 7 smaller layers, which we call IncepSE layers. There are three main design components in our model that leveraged its performance: the integration of a squeeze and excitation layer (SE) after the 1x1 convolutional “bottleneck” layer for the feature attention mechanism; additional convolutional layers and a skip connection in each IncepSE layer for better extraction of both local and global features; modifications of the last IncepSE layer to better suit the task of ECG analysis. For a clear illustration, the IncepSE model's architectural specifics are depicted in \autoref{fig:1}.

\subsection{Squeeze and Excitation Mechanism}
In each IncepSE layer, we integrate a Squeeze and Excitation layer (SE) \cite{SENet} right after the bottleneck convolutional layer for several key reasons. SE offers a self-attention mechanism to each channel, a feature widely acknowledged for its significant benefits in various network architectures \cite{sandler2019mobilenetv2}\cite{SE-ECG}. The bottleneck convolutional layer here is inspired by InceptionTime, which is a convolutional layer with both a kernel size and a stride of 1. InceptionTime's bottleneck is known for its ability to capture diverse and multi-scale features from time-series data when combined with large convolutions. Due to the channel reduction nature of the bottleneck, it can effectively incorporate these convolutions without overfitting by using too many parameters. By integrating SE, the bottleneck branch can further enhance the quality of these features by emphasizing the most informative channels. This undoubtedly leads to more discriminative representations, thereby allowing the model to focus on task-specific features and further improving generalization and adaptability. Moreover, the bottleneck convolution lends itself exceptionally well to SE integration, as it remarkably reduces the computational resource requirements.

The SE layer here is built using two fully connected (linear) layers, a ReLU and a sigmoid layer to normalize the outputs to $(0,1)$ as the attention weights to multiply to each channel before passing through the multiple-sized convolutional layers as shown. The information flow through the bottleneck and SE layers is demonstrated in the following equations.
\begin{equation} \label{eq7}
    s^{*}_i = \textrm{\textit{Conv}}^{1}_{32}(Input)
\end{equation}
\begin{equation} \label{eq8}
    Output = Sigmoid(Linear(ReLU(Linear(s^{*}_i)))) \times s^{*}_i
\end{equation}

\subsection{IncepSE Layer}
Inspired by the multiple convolutional branch design of InceptionTime, the design of each IncepSE layer is shown in \autoref{fig:1}. As seen from the picture, three relatively large convolutions with lengths of 10, 20 and 40 are applied to the output of the SE-bottleneck layers to effectively extract diverse multi-level temporal features, which follows the aforementioned strength of dimension reduction. We also introduce a convolutional layer with a kernel size of 3 into the max-pooling branch of InceptionTime, effectively transforming this branch into a convolutional layer with a sliding window of 5, all without substantially increasing the parameter count. Additionally, we introduce a branch containing a convolutional layer with a kernel size of 3, serving as a skip connection to stabilize gradient flow, removing the need for skip connections between IncepSE layers. By concatenating these two smaller kernel sizes with the global features obtained by the larger kernel size convolutions, these branches jointly facilitate the extraction of local features, thus further allowing our model to capture critical patterns at various scales while maintaining computational efficiency, which aligns with the heart of the success of InceptionTime. The information flow within an IncepSE cell is demonstrated in the following equations. In these equations, $s_i$ represents as the input of the i-th layer; moreover, the superscripts indicate the kernel size, while the subscripts denote the total number of channels.

\begin{equation} \label{eq1}
    b_0 = SE(\textrm{\textit{Conv}}^{1}_{32}(s_i)) \\
\end{equation}
\begin{equation} \label{eq2}
    b_1 = \textrm{\textit{Conv}}^{3}_{32}(\textrm{\textit{MaxPool}}^{3}(s_i)) \\
\end{equation}
\begin{equation} \label{eq3}
    b_2 = \textrm{\textit{Conv}}^{9}_{32}(b_0) \\
\end{equation}
\begin{equation} \label{eq4}
    b_3 = \textrm{\textit{Conv}}^{19}_{32}(b_0) \\
\end{equation}
\begin{equation} \label{eq5}
    b_4 = \textrm{\textit{Conv}}^{39}_{32}(b_0) \\
\end{equation}
\begin{equation} \label{eq6}
    s_{i+1} = ReLU(BN(Concat(b_1, b_2, b_3, b_4))) \, + \, \textrm{\textit{Conv}}^{3}_{128}(s_i) \\
\end{equation}

\subsection{Task-specific Configuration}

To enhance adaptability to diverse tasks within the dataset, the final IncepSE layer has twice the number of channels with a downsampling ratio of two and  includes a dropout layer. Due to the variations in the complexity of different ECG tasks, a deeper network with more channels is highly necessary, enabling our model to explore more abstract and intricate features when adapting to a wide range of tasks in the dataset. With all these configurations, our IncepSE significantly outperformed baselines and previous state-of-the-art models across all of the tasks in PTB-XL, as demonstrated in \autoref{result}.

Detailed experimental configurations can be found in \autoref{experiment}; however, it is important to highlight that PTB-XL suffers from severe imbalance labels. This poses a significant challenge during training, primarily due to the instability of the evaluation metric (macro AUCROC). Furthermore, we also believe that this uncertainty also comes from gradient corruption, i.e., gradient explosion and vanishing. Consequently, we propose precautionary measures by applying gradient clipping and weight decay techniques to enhance the gradient's quality, essentially to avoid gradient issues, thereby ensuring training stability. This training technique is proved in \autoref{numerical result} to be substantially beneficial for the inherently disproportional labels in \cite{Strodthoff:2020Deep}.



\section{Experiment} \label{experiment}

\subsection{Dataset and Experimental Settings}\label{dataset and setting}

In this study, we tested our methods on the PTB-XL dataset \cite{wagner2020ptb} for assessment of our model. Containing 21,837 10-second long clinical 12-lead ECG records of from 18,885 patients, the dataset provides comprehensive insights into cardiac activity. The dataset includes 71 distinct ECG statements (All), which can be grouped into 44 diagnostic (Diag), 19 form (Form), and 12 rhythm statements (Rhythm), and each record can contain more than just one statement. For diagnostic statements, they can be organized into 5 superdiagnostic classes (Super) or 23 subdiagnostic classes (Sub). While both sampling rates of 100 Hz and 500 Hz for each record was provided by the dataset, we only use the 100 Hz version only following the benchmark paper \cite{Strodthoff:2020Deep}. The dataset was split into ten folds, with the tenth fold recommended as the test set and the remaining nine folds as training and validation sets. Following this, we used the first eight folds for training, the ninth fold for validation and the tenth fold for testing.

Additionally, ECGs primarily consist of signals within the frequency range of [1, 45 Hz] \cite{Xu_preprocessing}. Therefore, prior to inputting the data into the model, we apply a Butterworth bandpass filter to the raw ECGs. This approach has yielded slight performance gains by effectively reducing noise and mitigating baseline distortions associated with high and low frequencies, respectively.

Our experimentation is conducted on the PTB-XL dataset, encompassing various tasks. Within each experiment, we employ either the OneCycle or ReduceOnPlateau scheduler during training, albeit with distinct combinations of learning rates and epochs. However, a consistent approach involves the utilization of gradient clipping of 0.1 along with weight decay of $10^{-4}$ as mentioned to avoid gradient exploding and vanishing. Our idea is implemented with Pytorch so that the hyperparameters provided in \autoref{tab: config} refer to the parameters of the Pytorch API.

\begin{table*}[h!]
    \caption{Experiments' configurations}
    \label{tab: config}
    \centering
    \begin{tabular} {| c c c c c |}
    \hline
         Tasks & Scheduler and parameters & learning rate & Numeber of epochs & Dropout  \\
         \hline
         All & \begin{tabular}{c}OneCycle:\\epochs=13\end{tabular} & $1e^{-2}$ & 15 & 0.00 \\
         \hline
         Diag & \begin{tabular}{c}OneCycle:\\epochs=16\end{tabular} & $1e^{-2}$ & 18 & 0.05 \\
         \hline
         Sub & \begin{tabular}{c}ReduceOnPlateau:\\factor=0.3, patience=1\end{tabular} & $1e^{-3}$ & 15 & 0.09 \\
         \hline
         Super & \begin{tabular}{c}OneCycle:\\epochs=14\end{tabular} & $1e^{-2}$ & 15 & 0.11 \\
         \hline
         Form & \begin{tabular}{c}OneCycle:\\epochs=16\end{tabular} & $1e^{-2}$ & 25 & 0.1 \\
         \hline
         Rhythm & \begin{tabular}{c}OneCycle:\\epochs=16\end{tabular} & $1e^{-2}$ & 20 & 0.1 \\
         \hline         
    \end{tabular}
    
\end{table*}

\subsection{Metrics and Implementation Details} \label{implementation}
All of the tasks are evaluated by averaging macro AUROC scores of ten runs, which aligns with the benchmark's default metric \cite{Strodthoff:2020Deep}. We also conduct experiments on other state-of-the-art models if they do not follow the conventional evaluation; however, we still follow the setup of the original paper. Each model was trained for a number of epochs reported in \autoref{tab: config}; afterward, the state with the highest validation's AUROC score was retrieved to be further evaluated on the test fold. In \autoref{result}, we provide the outcomes of all of our experiments along with ablation studies on each component in IncepSE as well as the training setups \autoref{numerical result}.

\section{Results} \label{result}

\subsection{Comparison Benchmarks} \label{comparison}

The results \autoref{tab: compare IncepSE} show that our model reaches competitive results across all of the tasks provided in PTB-XL and consistently outperforms the original architecture of InceptionTime by a large margin. Perhaps the most significant improvement lies in the overall performance (task "All") which covers every label presented in the dataset. We surpass previous state-of-the-art \cite{selfsupervisedECG} in the category of training end-to-end supervised learning by roughly 0.06 AUROC score. Hence, this shows the capability of Inception-like architecture and channel attention mechanism \cite{SENet} in the field of ECG.  With the right setups that prevent overfitting and gradient corruption, we were able to push the boundaries of 1D-CNNs in this dataset.
\begin{table*}[h!]
    \caption{IncepSE and others on PTB-XL tasks, the best results are highlighted in bold. We reassess ST-CNN-GAP-5 with the mentioned metric while following their orginal setups, which only ran on "super" task. On the other hand, we report the performances of 4FC-2LSTM-2FC given by \cite{selfsupervisedECG}.}
    \label{tab: compare IncepSE}
    \centering
    \begin{tabular} {| c c c c c c c |}
    \hline
         Model & All & Diag & Sub & Super & Form & Rhythm \\
         \hline
         xResNet \cite{DBLP:journals/corr/abs-1812-01187} & .925 & \textbf{.937} & .929 & .928 & .896 & .957 \\
         \hline
         ResNet1d-Wang \cite{wang2017time} & .919 & .936 & .928 & .930 & .880 & .946 \\
         \hline
         FCN-Wang \cite{wang2017time} & .918 & .926 & .927 & .925 & .869 & .931 \\
         \hline
         InceptionTime \cite{DBLP:journals/corr/abs-1909-04939} & .925 & .931 & .930 & .921 & .899 & .953 \\
         \hline
         ST-CNN-GAP-5 \cite{STT-CNN5} & - & - & - & .9318 & - & - \\
         \hline
         4FC-2LSTM-2FC \cite{selfsupervisedECG} & .9318 & - & - & - & - & - \\
         \hline
         \textbf{IncepSE} & \textbf{.9380} & .9351 & \textbf{.9342} & \textbf{.9343} & \textbf{.9013} & \textbf{.9684} \\
         \hline
         
    \end{tabular}
    
\end{table*}
\subsection{Numerical results} \label{numerical result}
We evaluate the effectiveness of our model with and without the Squeeze and Excitation (SE) layer, along with the mentioned minor adjustments. However, it is worth noting that, except for SE, we merge those remaining changes into a single group because when standing alone, they do not yield any significant improvements as opposed to incorporating all of them.

We run an experiment with InceptionTime integrated with only the SE bottleneck shown in the following \autoref{tab: compare component}. To ensure a fair comparison, all configurations incorporated gradient clipping and weight decay. Consequently, we reassessed InceptionTime with this configuration. The metric chosen is "super-diagnosis" (Super) due to the consistency of the outcomes compared to other tasks. The means and standard deviations of ten runs are recorded.

\begin{table}[h!]
    \caption{Components comparison on Super-diagnosis}
    \label{tab: compare component}
    \centering
    \begin{tabular} {| c c c |}
    \hline
         Model & Test mean & Test deviation \\
         \hline
         InceptionTime & .9329 & $14.19e^{-4}$ \\
         \hline
         InceptionTime + SE & .9337 & $6.10e^{-4}$ \\
         \hline
         IncepSE & .9343 & $4.58e^{-4}$ \\
         \hline         
    \end{tabular}
    
\end{table}

As clearly shown, each component leverages the performance of the overall architecture while mitigating the instability of the unbalanced dataset. It is essential to highlight that by adopting the novel methods of gradient clipping and weight decay, the performance of InceptionTime is enhanced significantly compared to the result reported by \cite{Strodthoff:2020Deep} and in \autoref{tab: compare IncepSE}.

We conduct a more in-depth exploration of how the stability techniques mentioned earlier (gradient clipping and weight decay) impact the performance of IncepSE. These experiments are carried out using the "All" classification task, which encompasses all other label categories in the dataset. Same as before, our configurations align with the setups in \autoref{tab: config} apart from the stabilize techniques.

Firstly, it become evident that weight clipping have the most significant influence on our model's results. Consequently, we proceed to conduct multiple experiments using various clipping values, as outlined in \autoref{tab: compare clipping}. Secondly, the choices of learning rate decay value do not have any noticeable impact on the performance; hence, they are not included in the table. However, this section also provides an experiment on the effectivenes of combining both methods at the end of this section.

As demonstrated, the performance of our model benefits substantially from clipping gradients. We believe this occurred due to the gradient exploding from having too many branches and skip-connections, which resulted in vigorous gradient backward flow through the bottleneck of IncepSE and InceptionTime. This causes the model to overfit on validation data, thereby losing robustness and generality.

\begin{table}[h!]
    \caption{Clipping values comparison on All-task}
    \label{tab: compare clipping}
    \centering
    \begin{tabular} {| c c c |}
    \hline
         Clipping value & Test mean & Test deviation \\
         \hline
         no clip        & .9332     & $2.23e^{-3}$   \\
         \hline
         0.5            & .9360     & $3.44e^{-3}$   \\
         \hline
         0.3            & .9369     & $2.36e^{-3}$   \\
         \hline
         0.1            & .9380     & $3.36e^{-3}$   \\
         \hline         
    \end{tabular}
    
\end{table}
Lastly, we include a study on the influence of the training that integrated both stabilizing methods. The task "Super" is chosen to minimize the potential impact of uncertainty on the results. Three runs are shown: the first run encompass all the configurations mentioned earlier, the second run solely employ weight decay, and the last run do not utilize any of the methods. It shows that both techniques worked in harmony with each other and overall enhance the performances as well as mitigate the deviations \autoref{tab: compare stability}.

\begin{table}[h!]
    \caption{Stability methods comparison on Super-task}
    \label{tab: compare stability}
    \centering
    \begin{tabular} {| c c c |}
    \hline
         Methods & Test mean & Test deviation \\
         \hline
         Clipping + weight decay & .9343     & $7.76e^{-4}$   \\
         \hline
         Weight decay            & .9341     & $10.52e^{-4}$   \\
         \hline
         None                    & .9337     & $11.60e^{-4}$   \\
         \hline        
    \end{tabular}
\end{table}

\section{Conclusion}
\subsection{Limitations and Future Directions}
Despite showing great performance, designing such networks tailored to specific tasks requires substantial effort and high levels of expertise, similar to the previous works. Moreover, the process of finetuning and trial-and-error is very time-consuming while still risking suboptimality. In the future, we aim to employ NAS (Neural Architecture Search) \cite{white2023neural} and other techniques to tackle this issue while further enhancing classification performance.

\subsection{Conclusion}
In this study, we proposed IncepSE, a 1D-CNN that capitalized on the strengths of both InceptionTime and channel attention mechanisms for the ECG classification task. Additionally, we introduced innovative training setups, incorporating stabilization techniques designed to address the challenges posed by the severely imbalanced PTB-XL dataset and gradient corruption. With these solutions, we achieved state-of-the-art performance on all classification tasks in this dataset. We hope that our findings not only contributed valuable insights to the construction of high-performing neural networks but also showed the importance of the training procedure in the task of ECG analysis, given the imbalance nature of ECG datasets.

\begin{acks}
This research was funded by Samsung R\&D Vietnam.
\end{acks}

\bibliographystyle{ACM-Reference-Format}
\bibliography{IncepSE.bib}

\end{document}